\documentclass[journal,twoside]{IEEEtran}

\usepackage[utf8]{inputenc} 
\usepackage[T1]{fontenc} 
\usepackage{amsmath}
\usepackage{amsfonts}
\usepackage{amsbsy}
\usepackage{amssymb}
\usepackage{bbm}
\usepackage[italicdiff]{physics}
\usepackage{proba}
\usepackage{enumerate}
\usepackage{cite}
\usepackage{wrapfig}
\usepackage{siunitx}
\usepackage{soul} 
\usepackage{times}
\usepackage{adjustbox}
\usepackage{xfrac}
\usepackage{url}
\usepackage{graphicx}
\usepackage{rotating}
\usepackage{mdframed}
\usepackage{cuted}
\usepackage{xcolor}

\mdfdefinestyle{fig}{skipabove=3pt,leftmargin=0pt,innerleftmargin=3pt,innertopmargin=3pt,skipbelow=3pt,rightmargin=0pt,innerrightmargin=3pt,innerbottommargin=3pt}

\newcommand\cS{{\mathcal{S}}}
\newcommand\las{{\sf{s}}}
\newcommand\ie{{{\it i.e.,~}}}
\newcommand\eg{{{\it e.g.,~}}}

\usepackage{xr-hyper}
\usepackage[hidelinks]{hyperref}

\title{\bf Adaptive Time-Resolved Mass Spectrometry \\with Nanomechanical Resonant Sensors}
 
\author{\vspace{1.34cm}{\large Alper~Demir}\\ \vspace{1cm}
{\small \textsf{Ko\c{c} University}}\\ \vspace{0.05cm}
{\small \textsf{Istanbul, Turkey}}\\ \vspace{0.1cm}
{\small \textsf{aldemir@ku.edu.tr}}\vspace{1cm}
}

\begin{document}

\maketitle
\flushbottom

\begin{abstract}
  Nanomechanical resonant sensors that are based on detecting and tracking the resonance frequency deviations due to events of interest are being advocated for a variety of applications. All sensor schemes currently in use are subject to a basic trade-off between accuracy and speed, while there is great interest in improving both in order to enable unprecedented and widespread applications. Based on a thorough understanding of the characteristics of current resonant sensor architectures, we propose adaptive and flexible sensor schemes. Unlike recently proposed time-resolved mechanical detection methods, the proposed schemes do not require ensemble averaging of the resonator response for many independent identical stimuli. Distinct one-time events can be detected in real-time with high time resolution with an accuracy that then improves considerably with elapsed time. While the proposed adaptive schemes also need to abide by the fundamental speed versus accuracy trade-off, we show that there is still ``some room at the bottom'' for improvement with sensor architecture innovations. Pareto optimal performance that reaches a bound that is imposed by the fundamental thermomechanical noise can be achieved.   
\end{abstract}

\section{Introduction}

\IEEEPARstart{N}{anomechanical} resonant sensors offer unprecedented sensitivities and are used in a variety of applications, such as mass spectrometry~\cite{chaste2012nanomechanical,naik2009towards,hanay2012single,Dominguez-Medina918,Ghadimi764} and atomic force microscopy ({\sc AFM})~\cite{albrecht1991frequency,kobayashi2009frequency,adams2016harnessing,baykara2015noncontact,giessibl2019qplus}. In mass spectrometry, the main sensing mechanism is based on tracking the resonance frequency deviations arising from the interactions of the nanomechanical resonator with the sample under study. 
Currently, three schemes are used for this purpose, namely the open-loop, feedback free ({\sc FF}) approach, and closed-loop schemes based on frequency-locked loops ({\sc FLL}) and self-sustaining oscillators ({\sc SSO}). These resonator tracking architectures were analyzed and the {accuracy} versus {speed} trade-off characteristics were discussed in detail in~\cite{demir2019fundamental,demir2020sensortradeoffs}.  
   
The speed-accuracy trade-off in the resonator tracking schemes arises from the characteristics of inherent noise sources affecting resonant sensors, \ie thermomechanical and detection noise. In order to increase the accuracy (sensitivity) of the sensor, one has to simply reduce the effective bandwidth of the system for better noise filtering. However, bandwidth reduction results in a larger response time. All of the standard resonator tracking schemes mentioned above have to endure this basic trade-off~\cite{demir2020sensortradeoffs}. On the other hand, noise reduction may also be achieved via {\em ensemble averaging}, as opposed to {\em temporal filtering}. In fact, recently proposed time-resolved mechanical motion detection techniques in atomic force microscopy~\cite{coffey2006time,giridharagopal2012submicrosecond,durmus2016,mascaro2019review} seem to be mostly based on some form of ensemble averaging of the resonator response for many identical stimuli. However, these techniques may be used only in cases where the events of interest have the form of multiple independent identical stimuli. They are not suitable for time-resolved detection of distinct one-time events that are typical in mass spectrometry applications.

The rigorous and thorough understanding that was developed in~\cite{demir2019fundamental,demir2020sensortradeoffs} for the fundamentals and trade-offs of standard resonator tracking schemes paves the way to the conception of adaptively accurate resonator tracking schemes. An adaptive scheme, in principle, would provide very fast but somewhat coarse initial estimates of frequency shifts, that are then improved continually with elapsed time after an event of interest occurs. While such an adaptive scheme also needs to abide by the basic trade-off between noise filtering and speed, we show that there is still ``some room at the bottom’’\cite{feynman1959plenty} for improvement, unraveled by the detailed analysis we present in this paper. The performance boost here is due to sensor architecture innovations alone, and is in addition to the improvement one would get by using higher quality resonators and better electronic or optical hardware. We propose an adaptively accurate, time-resolved resonator tracking scheme that can detect distinct, one-time events either in real-time operation or offline processing. We show that this Pareto optimal (with respect to speed and accuracy) scheme achieves a performance bound that is imposed by the unavoidable, fundamental thermomechanical noise of the resonator.   

We first discuss adaptively accurate resonator tracking schemes in Section~\ref{sec:adaptive} with a simple and intuitive perspective. We then propose a scheme based on adaptive Kalman filtering in Section~\ref{sec:kf} and rigorously compare it with the three standard schemes in Section~\ref{sec:compkfwothers}. Finally, we discuss practical implementation issues in Section~\ref{sec:akfpractical}.  

\section{Resonator tracking with adaptive speed versus accuracy characteristics}
\label{sec:adaptive}

The flexible speed versus accuracy characteristics of the {\sc FLL} scheme discussed in~\cite{demir2020sensortradeoffs} points to the possibility of constructing adaptively accurate resonator tracking schemes. Such an adaptive scheme would provide an {\em ultra fast response} immediately after an event of interest (causing a resonance frequency shift) with an estimate of the new value of the resonance frequency. It would then {\em continually} update this estimate, increasing its accuracy as time progresses following the event of interest. In principle, such a scheme could be implemented with an {\sc FLL} by adjusting the loop bandwidth (via the controller parameters) in an adaptive and automatic manner. This would involve real-time monitoring of the feedback error signal in the {\sc FLL}, and increasing the loop bandwidth dramatically if the error exceeds a certain threshold, indicating that an event has occurred. The loop would then respond very rapidly and lock to the shifted resonance. Subsequently, the loop bandwidth needs to be gradually reduced for better noise filtering and hence increased accuracy of tracking. There are some challenges in implementing such a scheme though, such as the conception of an effective and robust technique for detecting events based on monitoring the error signal and ensuring the stability of the feedback loop with time-varying controller parameters. An engineering solution for these challenges may be found, though with some effort and increased system complexity.

\begin{figure}
\centering
\includegraphics[width=0.9\columnwidth]{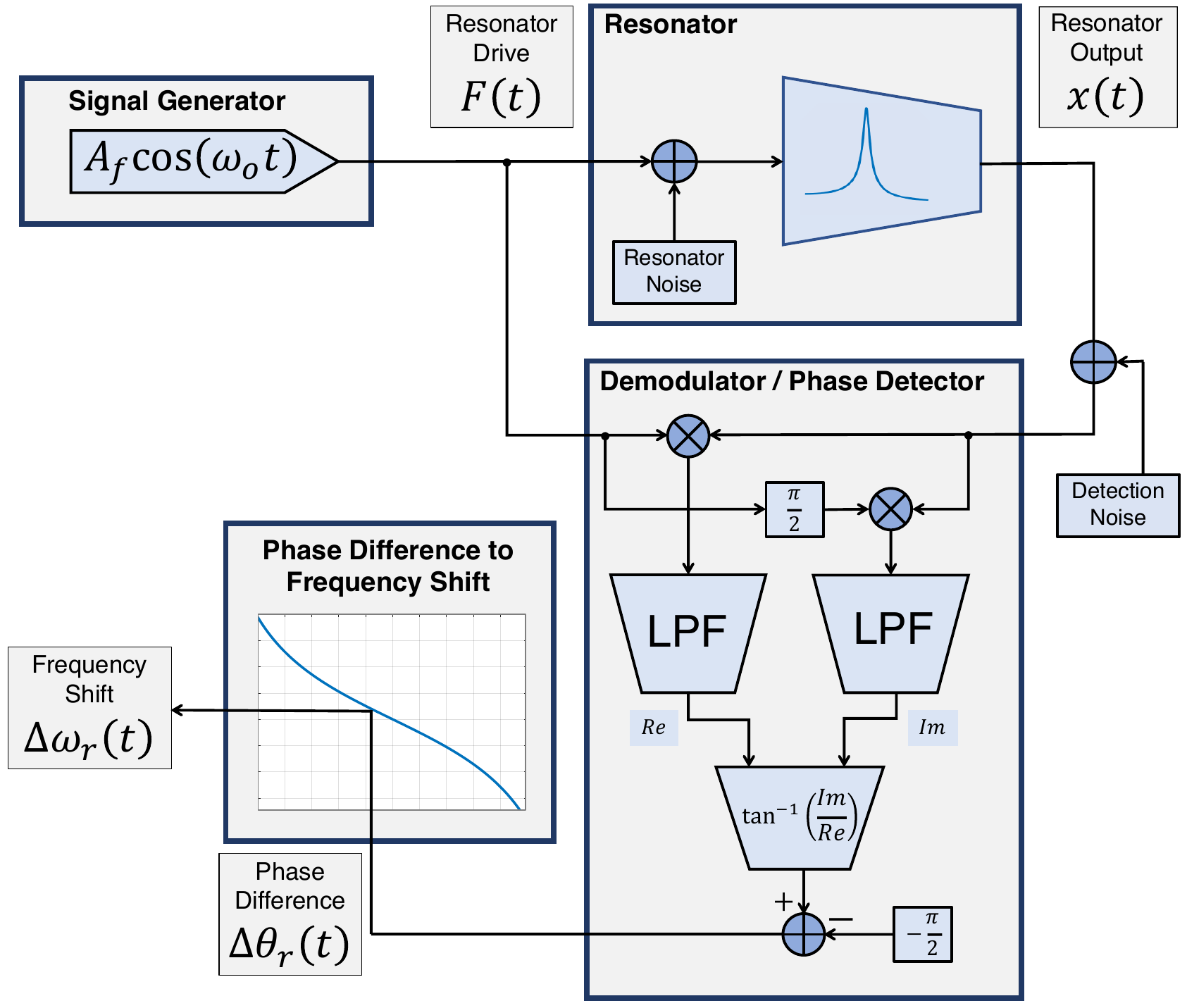}
\caption{Feedback-free (FF) resonator tracking scheme. {\sc LPF} is a low-pass filter with transfer function $H_\textsf{\tiny L}\pqty{\las}$. (Reproduced from \cite[Fig.~4]{demir2020sensortradeoffs}, Demir, Journal of Applied Physics 129, 044503 (2021), with the permission of AIP Publishing.)}
\label{fig:openloop}
\end{figure}
	
\begin{figure}
\centering
\includegraphics[width=0.8\columnwidth]{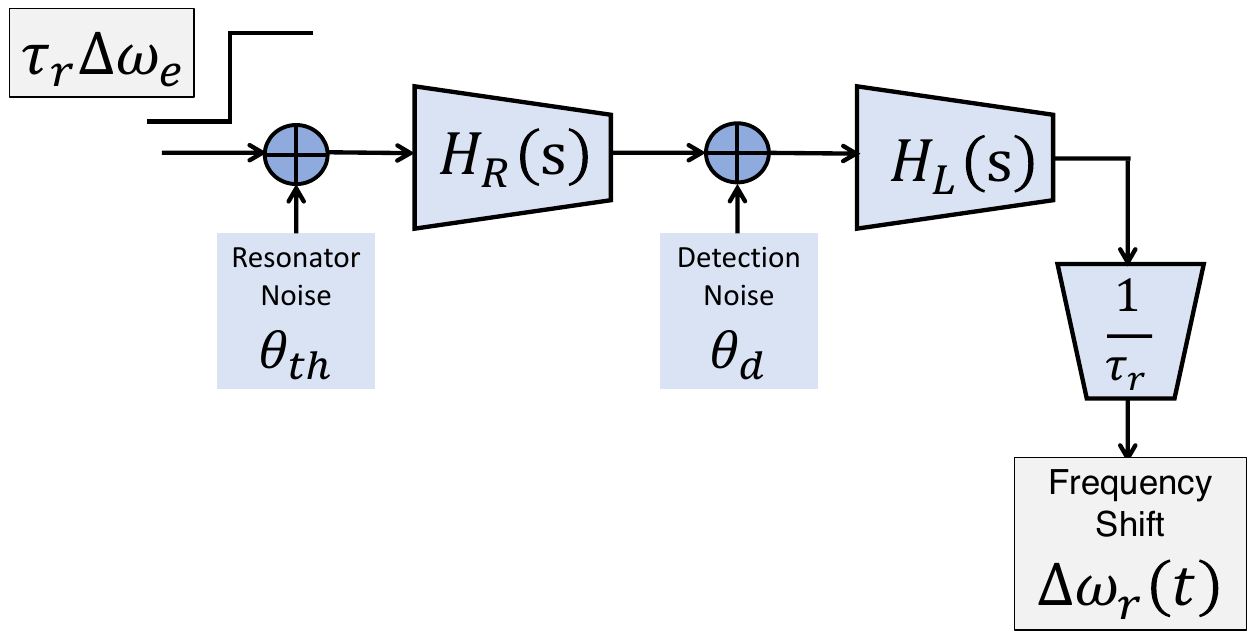}
\caption{Phase-domain model for {\sc FF} resonator tracking scheme.  $\Delta\omega_{\textsf{\tiny e}}$ represents a sudden resonance frequency shift (root cause) due to an event of interest. The frequency shift measured by the  {\sc FF} resonator scheme is denoted by $\Delta\omega_{\textsf{\tiny r}}\pqty{t}$. $H_\textsf{\tiny R}\pqty{\las}$ models the inherent resonator dynamics. $H_\textsf{\tiny L}\pqty{\las}$ represents the low-pass filtering in the demodulator. $\tau_{\textsf{\tiny r}}$ is the resonator response time constant.  (Reproduced from \cite[Fig.~7]{demir2020sensortradeoffs}, Demir, Journal of Applied Physics 129, 044503 (2021), with the permission of AIP Publishing.)} 
\label{fig:openloopphase}
\end{figure}

In the simple {\sc FF} scheme, shown in Fig.~\ref{fig:openloop}, the resonator is driven with a sinusoidal signal at a fixed frequency close to the resonance~\cite{demir2020sensortradeoffs}. The resonator response at the same frequency is demodulated (down-converted to baseband using the drive signal) in order to compute the phase shift with respect to the drive. This phase shift can then be used to infer the difference between the resonance frequency and the drive frequency based on the resonator characteristics. The {\sc FF} scheme does not offer a flexible speed versus accuracy characteristics as in the {\sc FLL}, and is limited by the inherent resonator characteristics~\cite{demir2020sensortradeoffs}. Thus, one may initially think that the {\sc FF} scheme is not amenable to adaptation as described above. In fact, by examining the very simple model that was developed in~\cite{demir2020sensortradeoffs} for the {\sc FF} scheme, shown in Fig.~\ref{fig:openloopphase}, one can conceive a detection scheme as follows. We assume that the response speed of the {\sc FF} scheme is limited by the resonator characteristics. That is,  with a high quality factor resonator, the linewidth of the resonator is much smaller than the bandwidth of the low-pass filter $H_\textsf{\tiny L}\pqty{\las}$ in the demodulator~\cite{sadeghi2020frequency}. Then, the response of the {\sc FF} scheme to an event that causes a sudden shift in the resonance frequency is simply determined by a very simple one-pole low-pass filter 
\begin{equation}
H_{\textsf{\tiny R}}\pqty{\las} =\frac{1}{1+\las\,\tau_{\textsf{\tiny r}}},
\label{eqn:restranfun}
\end{equation}
with a bandwidth equal to the resonator linewidth, where $\tau_{\textsf{\tiny r}}$ is the inherent resonator time constant~\cite{demir2020sensortradeoffs}. Thus, the step response of the {\sc FF} scheme to a sudden resonance frequency shift of 
$\Delta\omega_{\textsf{\tiny e}}$ at $t=0$ is simply given by 
\begin{equation}
      \Delta\omega_{\textsf{\tiny r}}\pqty{t} = \Delta\omega_{\textsf{\tiny e}}\pqty{1-e^{-t/\tau_{\textsf{\tiny r}}}}.
  \label{eqn:ffstepresponse}
\end{equation}  
Straightforward determination of the unknown $\Delta\omega_{\textsf{\tiny e}}$ based on such a transient response requires at least $t\approx\tau_{\textsf{\tiny r}}$ for the exponential factor to decay, hence it is quite slow. However, {\em prior knowledge} of the form of the response in \eqref{eqn:ffstepresponse} and $\tau_{\textsf{\tiny r}}$, {\em based on the simple but accurate model} in Fig.~\ref{fig:openloopphase}, enables one to determine $\Delta\omega_{\textsf{\tiny e}}$ immediately after event occurrence. One would simply measure the response $\Delta\omega_{\textsf{\tiny r}}\pqty{t_{m}}$ at any $t=t_{m}>0$ and evaluate 
\begin{equation}
      \Delta\omega_{\textsf{\tiny e}} = \frac{ \Delta\omega_{\textsf{\tiny r}}\pqty{t_{m}}}{\pqty{1-e^{-t_{m}/\tau_{\textsf{\tiny r}}}}}.
  \label{eqn:measfstepwithffstepresponse}
\end{equation}  
The {fast free time-resolved electrostatic force microscopy} ({\sc FF}-tr{\sc EFM})  scheme~\cite{giridharagopal2012submicrosecond} aiming to achieve better time resolution in {\sc AFM} is in principle 
similar to the one described above. 

The simplistic scheme represented by \eqref{eqn:measfstepwithffstepresponse} disregards noise and assumes that the event time ($t=0$, in this case) is known. Thus, there are two problems that need to be addressed. First, a scheme which can detect events very soon after they occur needs to be devised. Second, the naive measurement and evaluation in \eqref{eqn:measfstepwithffstepresponse} need to be replaced with an {\em optimal} estimation technique that takes both thermomechanical and detection noise into account. The second problem can be conveniently and properly addressed with the well established {\em Kalman Filter}~\cite{Kailath1968}, a least-squares based estimation technique for linear dynamical systems. The first problem can be addressed with the maximum-likelihood based technique described in~\cite{Willsky1976}, that augments a standard tracking-mode Kalman filter with a jump detection and estimation technique, turning it essentially into an adaptive filter. This adaptive Kalman filter is in principle similar to the adaptive {\sc FLL} scheme we described above, in the sense that 
the system features an automatically adjusted bandwidth for an initial rapid response and a subsequent increase in accuracy with time. The adaptive Kalman filter is founded on rigorous and well established least-squares and maximum-likelihood based estimation formulations, whereas the adaptive {\sc FLL} represents just a practical, engineering solution. 

Kalman filtering, with respect to both effectiveness and complexity considerations, critically relies on the availability of an accurate and simple dynamical model of the system, as well as prior stochastic characterizations of the noise sources~\cite{Kailath1968}. The model in Fig.~\ref{fig:openloopphase} for the {\sc FF} scheme, that was developed in~\cite{demir2020sensortradeoffs}, and the associated resonator and detection noise models are very accurate (especially for high $Q$ resonators), and as simple as it gets. Thus, we are well poised to develop a robust, effective and simple resonator tracking technique based on an adaptive Kalman filter. Kalman filtering based techniques have been previously considered in the {\sc AFM} literature~\cite{6913559,7118684,7480827,7908938,doi:10.1002/rnc.1025,doi:10.1063/1.1633963,salapaka2012atomic}. However, they were based on more complicated dynamical system models as compared with the one in Fig.~\ref{fig:openloopphase}, resulting in an implementation with higher complexity.  Furthermore, a rigorous analysis of the Kalman filtering based schemes with respect to speed versus accuracy characteristics, and performance comparisons with the traditional resonance frequency tracking architectures for mass spectrometry, seem to be lacking in the literature. 
                    
\section{Adaptive Kalman filtering based\\ tracking of resonators}
\label{sec:kf}
\subsection{Dynamical system model}
The adaptive Kalman filtering ({\sc AKF}) based resonator tracking scheme we propose is founded on the phase domain model in Fig.~\ref{fig:openloopphase}, which can be summarized as follows  (assuming $H_\textsf{\tiny R}\pqty{\las}$ representing the inherent resonator response has a bandwidth that is much smaller than the one for $H_\textsf{\tiny L}\pqty{\las}$ representing the demodulator filter)
\begin{equation}
  \begin{aligned}
      \tau_{\textsf{\tiny r}} \dv{}{t} {\Delta\theta_{\textsf{\tiny r}}\pqty{t}} + \Delta\theta_{\textsf{\tiny r}}\pqty{t} = \tau_{\textsf{\tiny r}}\Delta\omega_{\textsf{\tiny e}} + 
     \theta_{\textsf{\tiny th}}\pqty{t},\\
     \Delta\theta_{\textsf{\tiny ro}}\pqty{t} = \Delta\theta_{\textsf{\tiny r}}\pqty{t} + \theta_{\textsf{\tiny d}}\pqty{t},
  \end{aligned}
  \label{eqn:resonatorffphase}
\end{equation}
where $\Delta\theta_{\textsf{\tiny r}}\pqty{t}$ is the inherent (internal) phase deviation of the resonator that is influenced by thermomechanical noise $\theta_{\textsf{\tiny th}}$, and $\Delta\theta_{\textsf{\tiny ro}}\pqty{t}$ is the observed phase deviation that is corrupted by also detection noise $\theta_{\textsf{\tiny d}}$. 
Noting $\Delta\omega_{r}=\tfrac{\Delta\theta_{r}}{\tau_{\textsf{\tiny r}}}$ and $y_{\textsf{\tiny r}}=\tfrac{\Delta\omega_{\textsf{\tiny r}}}{\omega_{\textsf{\tiny r}}}$ thus $y_{\textsf{\tiny r}}=\tfrac{\Delta\theta_{\textsf{\tiny r}}}{\tau_{\textsf{\tiny r}}\,\omega_{\textsf{\tiny r}}}$, we divide both sides of \eqref{eqn:resonatorffphase} with $\tau_{\textsf{\tiny r}}\,\omega_{\textsf{\tiny r}}$ to obtain 
\begin{equation}
  \begin{aligned}
      \tau_{\textsf{\tiny r}} \dv{}{t} {y_{\textsf{\tiny r}}\pqty{t}} + y_{\textsf{\tiny r}}\pqty{t} = y_{\textsf{\tiny e}} + y_{\textsf{\tiny th}}\pqty{t},\\
     y_{\textsf{\tiny ro}}\pqty{t} = y_{\textsf{\tiny r}}\pqty{t} + y_{\textsf{\tiny d}}\pqty{t},
  \end{aligned}
  \label{eqn:resonatorfffracfreq}
\end{equation}
where all $y_{*}$ are fractional frequency quantities, $y_{\textsf{\tiny e}}$ is the root cause due to an event of interest, $y_{\textsf{\tiny r}}$ is the response of the resonator, the observable $y_{\textsf{\tiny ro}}$ includes also detection noise, and 
\begin{equation}
\begin{aligned}
y_{\textsf{\tiny th}}\pqty{t} = \frac{\theta_{\textsf{\tiny th}}\pqty{t}}{\tau_{\textsf{\tiny r}}\,\omega_{\textsf{\tiny r}}}\quad,\quad y_{\textsf{\tiny d}}\pqty{t} = \frac{\theta_{\textsf{\tiny d}}\pqty{t}}{\tau_{\textsf{\tiny r}}\,\omega_{\textsf{\tiny r}}}
\end{aligned}
\label{eqn:fracfreqnoise}
\end{equation}
are both approximated as white noise sources with (two-sided) spectral densities given by 
\begin{equation}
  \begin{aligned}
    \cS_{y_{\textsf{\tiny th}}}\pqty{\omega} &
      = \frac{k_{\textsf{\tiny B}}\,T}{m\,Q\,\omega_{\textsf{\tiny r}}^3\,A_{\textsf{\tiny rss}}^2},\\
    \cS_{y_{\textsf{\tiny d}}}\pqty{\omega} &= {\cal K}_{\textsf{\tiny d}}^2\:\cS_{y_{\textsf{\tiny th}}}\pqty{\omega},
  \end{aligned}
  \label{eqn:specynd}
\end{equation}
where 
$k_{\textsf{\tiny B}}$ is Boltzmann's constant, $T$ is temperature, $m$ is the mass of the resonator, $\omega_{\textsf{\tiny r}}$ is the resonance frequency, $Q$ is
the quality factor, and $A_{\textsf{\tiny rss}}$ is the vibrational amplitude of the resonator motion. The dimensionless factor ${\cal K}_{\textsf{\tiny d}}$ quantifies detection noise background with respect to thermomechanical noise, typically satisfying $0 < {\cal K}_{\textsf{\tiny d}} < 1$ when thermomechanical noise is resolved~\cite{demir2020sensortradeoffs}. 

We represent \eqref{eqn:resonatorfffracfreq} in state-space form and append a new state equation that captures the dynamics of $y_{\textsf{\tiny e}}$:  
\begin{equation}
  \begin{aligned}
  \dv{}{t} {y_{\textsf{\tiny e}}\pqty{t}} & = u_{\textsf{\tiny e}}\pqty{t},\\ 
      \dv{}{t} {y_{\textsf{\tiny r}}\pqty{t}} & =  \frac{y_{\textsf{\tiny e}}\pqty{t} - y_{\textsf{\tiny r}}\pqty{t}}{\tau_{\textsf{\tiny r}}} + \frac{y_{\textsf{\tiny th}}\pqty{t}}{\tau_{\textsf{\tiny r}}},\\
     y_{\textsf{\tiny ro}}\pqty{t} & = y_{\textsf{\tiny r}}\pqty{t} + y_{\textsf{\tiny d}}\pqty{t},
  \end{aligned}
  \label{eqn:resonatorfffracfreqss}
\end{equation}
where $u_{\textsf{\tiny e}}\pqty{t}$ characterizes the event. For instance, $u_{\textsf{\tiny e}}\pqty{t}=\Delta y\,\delta\pqty{t-t_{o}}$, with $\delta\pqty{t-t_{o}}$ as a time-shifted Dirac delta function, represents an instantaneous resonance frequency jump at $t=t_{o}$ creating a step in $y_{\textsf{\tiny e}}\pqty{t}$ and subsequently invoking a step response in  $y_{\textsf{\tiny r}}\pqty{t}$ and $y_{\textsf{\tiny ro}}\pqty{t}$. 
Thus, 
\begin{equation}
  \begin{aligned}
  \dv{}{t} 
  \begin{bmatrix} {y_{\textsf{\tiny e}}\pqty{t}} \\ {y_{\textsf{\tiny r}}\pqty{t}} \end{bmatrix} & = \begin{bmatrix} 0 & 0\\
  \frac{1}{\tau_{\textsf{\tiny r}}} & -\frac{1}{\tau_{\textsf{\tiny r}}} \end{bmatrix} \begin{bmatrix} {y_{\textsf{\tiny e}}\pqty{t}} \\ {y_{\textsf{\tiny r}}\pqty{t}} \end{bmatrix} +  \begin{bmatrix} u_{\textsf{\tiny e}}\pqty{t}\\ 0 \end{bmatrix} +\begin{bmatrix} 0 \\ \frac{y_{\textsf{\tiny th}}\pqty{t}}{\tau_{\textsf{\tiny r}}}\end{bmatrix}, \\
       y_{\textsf{\tiny ro}}\pqty{t} & = \begin{bmatrix} 0 & 1 \end{bmatrix} \begin{bmatrix} {y_{\textsf{\tiny e}}\pqty{t}} \\ {y_{\textsf{\tiny r}}\pqty{t}} \end{bmatrix} + y_{\textsf{\tiny d}}\pqty{t}.
  \end{aligned}
  \label{eqn:resonatorfffracfreqssmat}
\end{equation}
The implementation of the proposed scheme is best accomplished in the sampled-data domain. In order to lead to a digital implementation, we discretize time in \eqref{eqn:resonatorfffracfreqssmat} with a time-step of $h$ to obtain 
  \begin{align}
  \nonumber
  \begin{bmatrix} {y_{\textsf{\tiny e}}\bqty{k+1}} \\ {y_{\textsf{\tiny r}}\bqty{k+1}} \end{bmatrix} & = \underbrace{\begin{bmatrix} 1 & 0\\
  \frac{h}{\tau_{\textsf{\tiny r}}} & 1-\frac{h}{\tau_{\textsf{\tiny r}}} \end{bmatrix}}_{\mathbf F} \begin{bmatrix} {y_{\textsf{\tiny e}}\bqty{k}} \\ {y_{\textsf{\tiny r}}\bqty{k}} \end{bmatrix} +  \begin{bmatrix} u_{\textsf{\tiny e}}\bqty{k}\\ 0 \end{bmatrix} +\begin{bmatrix} 0 \\ \frac{y_{\textsf{\tiny th}}\bqty{k}}{\tau_{\textsf{\tiny r}}}\end{bmatrix}, \\
       y_{\textsf{\tiny ro}}\bqty{k} & = \underbrace{\begin{bmatrix} 0 & 1 \end{bmatrix}}_{\mathbf H} \begin{bmatrix} {y_{\textsf{\tiny e}}\bqty{k}} \\ {y_{\textsf{\tiny r}}\bqty{k}} \end{bmatrix} + y_{\textsf{\tiny d}}\bqty{k},
\label{eqn:resonatorfffracfreqssmatdisc}
\end{align}
with
\begin{equation}
\begin{aligned}
y_{\textsf{\tiny th}}\bqty{k} \sim \mathcal{N}\pqty{0, h\,\cS_{y_{\textsf{\tiny th}}}},\\
y_{\textsf{\tiny d}}\bqty{k} \sim \mathcal{N}\pqty{0, \textsf{\small BW}_{\textsf{\tiny L}}\,\cS_{y_{\textsf{\tiny d}}}}, 
\end{aligned}
\end{equation}
where $\mathcal{N}\pqty{m, v}$ represents a Gaussian random variable with mean $m$ and variance $v$. $u_{\textsf{\tiny e}}\bqty{k}=\Delta y\,\delta\bqty{k-k_{o}}$,  with $\delta\bqty{k-k_{o}}$ as a time-shifted Kronecker delta function, represents an instantaneous resonance frequency jump at $k=k_{o}$. 
$\textsf{\small BW}_{\textsf{\tiny L}}$ is the detection bandwidth (noise equivalent bandwidth\footnote{Defined as the cut-off frequency of an ideal low-pass filter which passes the same amount of noise power as the actual filter, equal to $\tfrac{\pi}{2}\omega_{\textsf{3dB}}$ for a first-order filter where $\omega_{\textsf{3dB}}$ is its  (one-sided) 3-dB bandwidth. Hence, $\textsf{\small BW}_{\textsf{\tiny L}}=\tfrac{2}{2\pi}\tfrac{\pi}{2}\omega_{\textsf{3dB}} =\tfrac{\omega_{\textsf{3dB}}}{2}$, since $\textsf{\small BW}_{\textsf{\tiny L}}$ is two-sided and expressed in $Hz$.}, two-sided, expressed in {\it Hz}) of the low-pass filter $H_\textsf{\tiny L}\pqty{\las}$ in the {\sc IQ} demodulator, typically set to be wider than the linewidth of the resonator, but limited by the resonance frequency $\omega_{r}$.  $y_{\textsf{\tiny th}}\bqty{k}$ and $y_{\textsf{\tiny d}}\bqty{k}$ are assumed to be uncorrelated with each other, and also in time \ie they are both discrete-time white noise processes. We define
\begin{equation}
{\mathbf Q} =  \begin{bmatrix}0 & 0 \\0 & \tfrac{h\,\cS_{y_{\textsf{\tiny th}}}}{\tau_{\textsf{\tiny r}}^{2}} \end{bmatrix},\quad {\mathbf R} =  \begin{bmatrix} \textsf{BW}_{\textsf{\tiny L}}\,\cS_{y_{\textsf{\tiny d}}}\end{bmatrix}
\end{equation}
as covariance matrices. Kalman filter relies on the simple, discrete-time model of the resonator 
in \eqref{eqn:resonatorfffracfreqssmatdisc} in order to make much faster (and optimal) predictions than otherwise allowed by the slow resonator response. 
\subsection{Tracking mode Kalman filter}
The recursive prediction and update equations for the tracking mode Kalman filter~\cite{Kailath1968} are as follows:
\begin{flalign}
\nonumber
&\text{\bf Predict}&\\
&\quad\quad\mathbf{\hat{y}}\bqty{k|k-1} = \mathbf{F}\,\mathbf{\hat{y}}\bqty{k-1|k-1},&\\
&\quad\quad\mathbf{C}\bqty{k|k-1} =  \mathbf{F}\,\mathbf{C}\bqty{k-1|k-1}\,\mathbf{F}^{T}+\mathbf{Q},&\\
\nonumber
&\text{\bf Observe}&\\
\label{eqn:kalmaninnovation}
&\quad\quad\mathbf{\tilde{y}}\bqty{k} =  y_{\textsf{\tiny ro}}\bqty{k}-\mathbf{H}\,\mathbf{\hat{y}}\bqty{k|k-1},&\\
&\quad\quad\mathbf{V}\bqty{k} =  \mathbf{H}\,\mathbf{C}\bqty{k|k-1}\,\mathbf{H}^{T}+\mathbf{R},&\\
\nonumber
&\text{\bf Kalman gain}&\\
\label{eqn:kalmangain}
&\quad\quad\mathbf{K}\bqty{k} = \mathbf{C}\bqty{k|k-1}\,\mathbf{H}^{T}\,\mathbf{V}^{-1},&\\
\nonumber
&\text{\bf Estimate}\\
&\quad\quad\mathbf{\hat{y}}\bqty{k|k} = \mathbf{\hat{y}}\bqty{k|k-1} + \mathbf{K}\bqty{k}\,\mathbf{\tilde{y}}\bqty{k},&\\
\nonumber
&\text{\bf Estimate covariance}&\\
&\quad\quad\mathbf{C}\bqty{k|k} = \pqty{\mathbf{I}-\mathbf{K}\bqty{k}\,\mathbf{H}}\,\mathbf{C}\bqty{k|k-1},& 
\end{flalign} 
where 
\begin{equation}
\mathbf{y} = \begin{bmatrix} {y_{\textsf{\tiny e}}} \\ {y_{\textsf{\tiny r}}} \end{bmatrix} 
\end{equation}
is the system state vector, $\mathbf{\tilde{y}}$ is the difference between the actual observation and the model based prediction for the observable resonator output $y_{\textsf{\tiny ro}}$, the $\hat{\bullet}\bqty{n|m}$ notation represents the estimate of a quantity $\bullet$ at time $n$ given the observations for up to time $m\leq n$, $\mathbf{K}$ is the Kalman gain, and $\mathbf{C}$ is the $2\times 2$ covariance matrix for the state estimate. The first diagonal entry of $\mathbf{C}$ quantifies the accuracy of the estimate for the root cause $y_{\textsf{\tiny e}}$ for the resonance frequency. 

In tracking mode, $y_{\textsf{\tiny e}}$ is time-invariant assuming no events occur, with $u_{\textsf{\tiny e}}=0$. In this case, the Kalman filter recursively and continually updates its estimate for $y_{\textsf{\tiny e}}$. The recursions for $\hat{y}_{\textsf{\tiny e}}$ and the covariance matrix $\mathbf{C}$ converge to a fixed point. The steady-state error variance for the estimate $\hat{y}_{\textsf{\tiny e}}$ at this fixed point is in fact zero, meaning that the estimate is perfect. However, this perfect estimate is, in principle, achieved only at convergence, \ie given infinite time. Intuitively, one can interpret the steady-state Kalman filter (producing a perfect estimate) as a system with {\em zero noise bandwidth}, \ie all of the noise is being filtered but at the same time the system has lost its ability to respond to any changes in the estimated quantity. From a practical point of view, the accuracy of the estimates produced by the Kalman filter needs to be assessed in a transient manner immediately following event occurrence. Before we can perform this characterization, we first need to introduce the event detection and estimation technique~\cite{Willsky1976} that will augment the Kalman filter.  
\subsection{Event detection and Kalman filter adaptation}
A sudden resonance frequency shift of an {\em unknown size} $\Delta y$ at an {\em unknown time} $k_{o}$, due to an event of interest, will manifest itself with $u_{\textsf{\tiny e}}\bqty{k}=\Delta y\,\delta\bqty{k-k_{o}}$ in \eqref{eqn:resonatorfffracfreqssmatdisc}. The event detection technique proposed in~\cite{Willsky1976} that we adopt is founded on a rigorous maximum-likelihood based estimation formulation, leading to an efficient implementation based on recursive prediction and update equations as in the tracking mode Kalman filter. We provide here an outline, and a somewhat imprecise but intuitive description for the technique, and refer the reader to~\cite{Willsky1976} for the details. 

The difference between the actual observation and the model based prediction for the observable resonator output $y_{\textsf{\tiny ro}}$ in \eqref{eqn:kalmaninnovation}, known as the residual or the innovation~\cite{Kailath1968} in a Kalman filter, is continually monitored via recursive equations. At every point in a prescribed sliding time window, the residuals are {\em projected} onto {\em sudden resonance frequency shift event signatures} with sliding occurrence times within the window. The event time is determined based on a maximum-likelihood rule when the residual projections exceed a certain threshold. If an event is detected as such at a certain estimated time instant $\hat{k}_{o}$, a corresponding maximum-likelihood based estimate for the size $\Delta\hat{y}$ of the frequency shift is also computed. The accuracy of this estimate is quantified via the covariance (matrix) as in the tracking mode Kalman filter. Subsequent to event detection, the Kalman filter in tracking mode is reset by updating both the state estimate $\mathbf{\hat{y}}$ and the corresponding estimate covariance matrix $\mathbf{C}$. The frequency shift estimate $\Delta\hat{y}$ at event detection is associated with a larger variance, since there is very little effective noise filtering. As a result, the Kalman gain $\mathbf{K}$ in \eqref{eqn:kalmangain} increases dramatically after the reset following event detection. This essentially corresponds to widening the bandwidth of the system, leading to a much faster response but also much less noise filtering. The Kalman filter then recursively and continually updates the estimate $\Delta\hat{y}$, with the variance of this estimate and hence the Kalman gain continually decreasing in the meantime. Thus, the effective bandwidth of the Kalman filter is continually reduced after the reset, eventually ending up at the steady-state situation described before where the estimate becomes perfect with zero variance. In a practical setting, this detection scheme is to be used when multiple events occur with some time gap between them. In this case, the initial variance (quantifying accuracy) of the frequency shift estimate right after detection and how fast it decreases with time is the key. 

The recursive equations for the detection of frequency shift events, and the maximum-likelihood estimators for the event time $\hat{k}_{o}$ and the size $\Delta\hat{y}$ of the frequency shift are as follows~\cite{Willsky1976}:
\begin{flalign}
\nonumber
&\text{\bf{for}~}{k-M<m<k}:&\\
\nonumber
&\quad\quad\mathbf{G}\bqty{k;m} = \mathbf{H}\,\bqty{\mathbf{F}^{k-m}-\mathbf{F}\,\mathbf{L}\bqty{k-1;m}},&\\
\nonumber
&\quad\quad\mathbf{CI}\bqty{k;m} = \mathbf{G}^{T}\bqty{k;m}\,\mathbf{V}^{-1}\,\mathbf{G}\bqty{k;m}+\mathbf{CI}\bqty{k-1;m},&\\
\nonumber
&\quad\quad\mathbf{d}\bqty{k;m} = \mathbf{G}^{T}\bqty{k;m}\,\mathbf{V}^{-1}\,\mathbf{\tilde{y}}\bqty{k}+\mathbf{d}\bqty{k-1;m},&\\
\nonumber
&\quad\quad\mathbf{L}\bqty{k;m} = \mathbf{K}\bqty{k}\,\mathbf{G}\bqty{k;m}+\mathbf{F}\,\mathbf{L}\bqty{k-1;m},&\\
\nonumber 
&\quad\quad\mathbf{a}\bqty{k;m} = \begin{bmatrix} 1 & 0 \end{bmatrix}\,\mathbf{CI}\bqty{k;m}\, \begin{bmatrix} 1 & 0 \end{bmatrix}^{{T}}\\
\nonumber 
&\quad\quad\mathbf{b}\bqty{k;m} = \begin{bmatrix} 1 & 0 \end{bmatrix}\,\mathbf{d}\bqty{k;m}\\
&\quad\quad\mathbf{l}\bqty{k;m} = \frac{\mathbf{b}^{2}\bqty{k;m}}{\mathbf{a}\bqty{k;m}}
\label{eqn:eventdetection}
\\ 
\nonumber
&\text{\bf{at}~}{k}:&\\
\nonumber
&\quad\quad\mathbf{CI}\bqty{k;k} = \mathbf{H}^{T}\,\mathbf{V}^{-1}\,\mathbf{H},&\\
\nonumber
&\quad\quad\mathbf{d}\bqty{k;k} = \mathbf{H}^{T}\,\mathbf{V}^{-1}\,\mathbf{\tilde{y}}\bqty{k},&\\
\nonumber
&\quad\quad\mathbf{L}\bqty{k;k} = \mathbf{K}\bqty{k}\,\mathbf{H}&
\end{flalign} 
The quantities above are recursively computed in a window of length $M$ as above, where $k$ is the current time and $m$ indicates a time point in the window starting at $k$ and going back to $k-M$. When $\mathbf{l}\bqty{k;m} >\textbf{threshold}$, an event is detected and the estimates for the event time $\hat{k}_{o}$ and the size $\Delta\hat{y}$ of the frequency shift are computed as follows. $\hat{k}_{o}$ is the value of $m$ that maximizes $\mathbf{l}\bqty{k;m}$, and 
\begin{equation}
\Delta\hat{y}\bqty{k} = \frac{\mathbf{b}\bqty{k;\hat{k}_{o}}}{\mathbf{a}\bqty{k;\hat{k}_{o}}}.
\end{equation}
Here, $k-\hat{k}_{o}$ is the detection delay time~\cite{Willsky1976}, and it can be controlled by adjusting the \textbf{threshold}, with lower values corresponding to shorter delays. However, making \textbf{threshold} too small may result in false alarms. 
Once an event is detected as such, the Kalman filter is reset by updating both the state estimate $\mathbf{\hat{y}}$ and the corresponding estimate covariance matrix $\mathbf{C}$ as follows
\begin{align}
\nonumber
\mathbf{\hat{y}}\bqty{k|k}_{\text{new}}=&\,\mathbf{\hat{y}}\bqty{k|k}_{\text{old}} + \Bqty{\mathbf{F}^{k-\hat{k}_{o}}-\mathbf{L}\bqty{k;\hat{k}_{o}}} \Delta\hat{y}\bqty{k} \begin{bmatrix} 1 & 0 \end{bmatrix}^{{T}},\\ \nonumber
\mathbf{C}\bqty{k|k}_{\text{new}} =&\,\mathbf{C}\bqty{k|k}_{\text{old}} \\ 
+&\Bqty{\mathbf{F}^{k-\hat{k}_{o}}-\mathbf{L}\bqty{k;\hat{k}_{o}}}\,\mathbf{CI}^{-1}\bqty{k;\hat{k}_{o}} 
\\&\Bqty{\mathbf{F}^{k-\hat{k}_{o}}-\mathbf{L}\bqty{k;\hat{k}_{o}}}.
\nonumber
\end{align}
The variance of the initial estimate for  $\Delta\hat{y}$ is given by~\cite{Willsky1976}:
\begin{equation}
{\text{\bf var}}\pqty{\Delta\hat{y}} = \begin{bmatrix} 1 & 0 \end{bmatrix}\,\mathbf{CI}^{-1}\bqty{k;\hat{k}_{o}}\, \begin{bmatrix} 1 & 0 \end{bmatrix}^{{T}}.
\end{equation}
\subsection{Event frequency shift estimation accuracy}
 We next consider the transient behavior of the frequency shift estimate accuracy as a function of elapsed time following event detection. The analytical quantification of the estimate accuracy is rather challenging, due to the recursive nature of the predict and update equations of the Kalman filter as well as the equations in \eqref{eqn:eventdetection} for event detection and estimation~\cite{Willsky1976}. However, the difficulty of the problem at hand is alleviated due to the very simple nature of the dynamical system model in \eqref{eqn:resonatorfffracfreqssmatdisc} that we use. Still, derivation of a precise expression for the transient behavior of the estimate variance is not tractable. However, by employing reasonable and justifiable approximations, we were able to derive an approximate analytical expression that matches both the numerically computed results based on an evaluation of the recursive equations above, as well as the ones obtained from extensive Monte Carlo simulations. An analytical quantification of the accuracy versus speed characteristics of the Kalman filtering based technique is needed for one-to-one comparisons with the traditional resonator tracking schemes. By performing considerable and tedious algebra with the aid of a symbolic computer algebra package, we were able to derive the following expression for the variance of the frequency shift estimate $\Delta\hat{y}$
\begin{equation}
{\text{\bf var}}\pqty{\Delta\hat{y}} = \frac{\begin{array}{c}\textsf{Z}+ \sqrt{\cS_{y_{\textsf{\tiny th}}}t_{e}\:\textsf{Z}}\end{array}} {2\,t_{e}^{2}}
\label{eqn:varAKF}
\end{equation} 
with
\begin{equation}
\textsf{Z} = {\cS_{y_{\textsf{\tiny th}}}t_{e}+4\,\textsf{\small BW}_{\textsf{\tiny L}}\,\cS_{y_{\textsf{\tiny d}}} \tau_{\textsf{\tiny r}}^{2}}
\end{equation}
where $t_{e}$ is the elapsed time since the event. 
The small--$t_{e}$ asymptote for this expression is given by 
\begin{equation}
{\text{\bf var}}\pqty{\Delta\hat{y}} = {2\,\textsf{\small BW}_{\textsf{\tiny L}}\,\cS_{y_{\textsf{\tiny d}}}} \pqty{\frac{\tau_{\textsf{\tiny r}}}{t_{e}}}^{2}
\end{equation} 
and for the large--$t_{e}$ asymptote, we have  
\begin{equation}
{\text{\bf var}}\pqty{\Delta\hat{y}} = \frac{\cS_{y_{\textsf{\tiny th}}}}{t_{e}}
\label{eqn:AKFvarhiasymp}
\end{equation} 
The two asymptotes intersect at the point 
\begin{equation}
t_{e}^{*} =  {2\,\textsf{\small BW}_{\textsf{\tiny L}}\,\tau_{\textsf{\tiny r}}^{2}
\,\frac{\cS_{y_{\textsf{\tiny d}}}}{\cS_{y_{\textsf{\tiny th}}}}} = 2\,\textsf{\small BW}_{\textsf{\tiny L}}\tau_{\textsf{\tiny r}}^{2}
\,{\cal K}_{\textsf{\tiny d}}^2
\label{eqn:asympintersectAKF}
\end{equation}
For small values of $t_{e}$, detection noise limits the accuracy of the estimate, since the bandwidth of the system is initially large. As $t_{e}$ becomes larger, the effective bandwidth of the system is reduced continually, filtering more and more of detection noise. The estimate accuracy eventually becomes limited by the thermomechanical noise of the resonator. 

\section{Comparison of \\Speed versus Accuracy characteristics} 
\label{sec:compkfwothers}

We quantify the quality of the estimate $\Delta\hat{y}$, for a resonance frequency shift of $\Delta y$, provided by various resonator tracking schemes in terms of their {\em Mean Squared Error} (\textsf{\small MSE}) (or with root mean squared error \textsf{\small RMSE} as its square root), defined by 
\begin{align}
\nonumber
\textsf{\small MSE} &= \text{\bf E}\bqty{\pqty{\Delta\hat{y}-\Delta y}^{2}} = \text{\bf var}\pqty{\Delta\hat{y}-\Delta y} + \pqty{\vb{E}\bqty{\Delta\hat{y}-\Delta y}}^{2}\\
&= \text{\bf var}\pqty{\Delta\hat{y}} + \pqty{\text{\bf bias}}^{2}
\label{eqn:mseakf}
\end{align}    
where where $\text{\bf E}\bqty{\cdot}$ denotes the probabilistic expectation operator. 
Thus, \textsf{\small MSE} is the sum of two terms, the variance and the square of the bias of the estimator. In general, both the variance and the bias are functions of the elapsed time $t_{e}$ after an event. 

All three of the standard {\sc FF}, {\sc FLL} and {\sc SSO} schemes provide {\em biased} estimates for $\Delta y$~\cite{demir2020sensortradeoffs}. The biases of these schemes can be directly determined using the {\it frequency step response} ({\sc FSTR}) as follows 
\begin{equation}
\text{\bf bias}\pqty{t_{e}} =  \Delta y\,\bqty{1-f_{\textsf{\tiny str}}\pqty{t_{e}}}
\end{equation} 
where $f_{\textsf{\tiny str}}\pqty{t_{e}}$ is normalized
so that $f_{\textsf{\tiny str}}\pqty{0}=0$ and $f_{\textsf{\tiny str}}\pqty{t_{e}\rightarrow\infty} =
1$~\cite{demir2020sensortradeoffs}. Thus, the biases of all three schemes are equal to the unknown frequency shift $\Delta y$ immediately following the event. The estimators become (asymptotically) unbiased as $t_{e}\rightarrow\infty$. 
In order to arrive at simple analytical expressions, we will approximate the {\sc FSTR} of the three schemes as follows
\begin{equation}
\begin{aligned}
 f^{\textsf{\tiny FF}}_{\textsf{\tiny str}}\pqty{t_{e}} & = 1-e^{-t/\tau_{\textsf{\tiny r}}}\\
 f^{\textsf{\tiny FLL}}_{\textsf{\tiny str}}\pqty{t_{e}} = 1-e^{-t/\tau_{\textsf{\tiny FLL}}}\:\:,&\:\:
 f^{\textsf{\tiny SSO}}_{\textsf{\tiny str}}\pqty{t_{e}} = 1-e^{-t/\tau_{\textsf{\tiny SSO}}}
\end{aligned}
\end{equation}
where 
\begin{equation}
\tau_{\textsf{\tiny FLL}} = \tau_{\textsf{\tiny SSO}} = \frac{1}{\omega_{\textsf{3dB}}}= \frac{1}{\textsf{2\,BW}_{\textsf{\tiny FLL}}} = \frac{1}{\textsf{2\,BW}_{\textsf{\tiny SSO}}}
\end{equation} 
is the time-constant for the effective low-pass filter (assumed first-order for the above) that models the detection/demodulator bandwidth for the {\sc FLL} and {\sc SSO} schemes, and $\textsf{BW}_{\textsf{\tiny FLL}}=\textsf{BW}_{\textsf{\tiny SSO}}$ is its two-sided noise equivalent bandwidth expressed in $Hz$. For the {
\sc FLL} scheme, we assume that the loop bandwidth is chosen to be the same as the bandwidth of the filter in the demodulator~\cite{demir2020sensortradeoffs}.    

The variances of the estimators for the {\sc FF}, {\sc FLL} and {\sc SSO} schemes can be computed by simply integrating the spectral density of fractional frequency fluctuations, $\cS_{y_{\textsf{\tiny r}}}\pqty{\omega}$, over the system bandwidth. We do this computation by multiplying $\cS_{y_{\textsf{\tiny r}}}\pqty{\omega=0}$ in \cite[eqn.~72]{demir2020sensortradeoffs} with the corresponding noise equivalent bandwidth as follows
  \begin{align}
    \nonumber
    \text{\bf var}^{\textsf{\tiny FF}}\pqty{\Delta\hat{y}} & = \cS_{y_{\textsf{\tiny r}}}^{\textsf{\tiny FF}}\pqty{0}\,\frac{\omega_{r}}{4 Q} \\ \nonumber & = \cS_{y_{\textsf{\tiny th}}}\,\frac{\omega_{r}}{4 Q}\,\pqty{1+{\cal
        K}_{\textsf{\tiny d}}^2} \\
        \label{eqn:varFFFLLSSO}
    \text{\bf var}^{\textsf{\tiny FLL}}\pqty{\Delta\hat{y}} & = \cS_{y_{\textsf{\tiny r}}}^{\textsf{\tiny FLL}}\pqty{0}\,\textsf{BW}_{\textsf{\tiny FLL}}\\ \nonumber & = \cS_{y_{\textsf{\tiny th}}}\,\textsf{BW}_{\textsf{\tiny FLL}}\,\pqty{1+{\cal
        K}_{\textsf{\tiny d}}^2}\\
        \nonumber
 \text{\bf var}^{\textsf{\tiny SSO}}\pqty{\Delta\hat{y}} & = \cS_{y_{\textsf{\tiny r}}}^{\textsf{\tiny SSO}}\pqty{0}\,\textsf{BW}_{\textsf{\tiny SSO}}\\ \nonumber & = \cS_{y_{\textsf{\tiny th}}}\,\textsf{BW}_{\textsf{\tiny SSO}}\,
 \pqty{1+{\cal
        K}_{\textsf{\tiny d}}^2}
    \nonumber
  \end{align}  
where $\cS_{y_{\textsf{\tiny th}}}$ is as given in \eqref{eqn:specynd}, and $\frac{\omega_{r}}{4 Q}$ is the two-sided noise equivalent bandwidth (in $Hz$) for the resonator assuming a first-order Lorentzian line shape. The (steady-state) variances above are all independent of $t_{e}$. The {\sc FF}, {\sc FLL} and {\sc SSO} schemes are  non-adaptive, with a fixed noise filtering bandwidth.  

The {\sc AKF} based scheme provides an {\em unbiased} estimator, with {\em zero} bias for all $t_{e}$. On the other hand, due to its adaptive noise filtering bandwidth, the variance of the estimator, given in  \eqref{eqn:varAKF}, is dependent on $t_{e}$.  

For {\sc FF}, {\sc FLL}, and {\sc SSO},  \textsf{\small MSE} is finite at $t_{e}=0$ and asymptotes to a nonzero value as $t_{e}\rightarrow\infty$. \textsf{\small MSE} depends on $\Delta y$, the frequency shift to be estimated, in fact dominated by it for small $t_{e}$, but asymptotes to a value that is independent of it. For both {\sc FLL} and {\sc SSO}, estimate variance is proportional to the detection bandwidth $\textsf{BW}_{\textsf{\tiny FLL}} = \textsf{BW}_{\textsf{\tiny SSO}}$. On the other hand, the amount of bias decreases as detection bandwidth is made larger. Thus, for a given $t_{e}$, there is an optimal value for the detection bandwidth.   

For {\sc AKF}, \textsf{\small MSE} is infinite at $t_{e}=0$ and asymptotes to zero at $t_{e}\rightarrow\infty$. \textsf{\small MSE} is independent of $\Delta y$ for all $t_{e}$. However, it does not make sense to use this scheme at very small $t_{e}$ where \textsf{\small RMSE} is on the order of $\Delta y$ (or larger) to be detected. Thus, even though \textsf{\small MSE} is independent of $\Delta y$, $\Delta y$ determines the smallest $t_{e}$ at which it makes sense to use the {\sc AKF} estimate. Using {\sc AKF} at a very small $t_{e}$ as such will make the reliable detection of event times impossible or non-robust. For small $t_{e}$, \textsf{\small MSE} is proportional to $\textsf{BW}_{\textsf{\tiny L}}$, and for large $t_{e}$, it is independent of it. Thus, $\textsf{BW}_{\textsf{\tiny L}}$ can be made as small as the linewidth of the resonator, or a little larger, so that the inherent response of the resonator is not modified by the filter. Finally, it should be noted that \textsf{\small MSE} for $t_{e}\rightarrow 0$ and $t_{e}\rightarrow\infty$ are obviously not meaningful from a practical point of view. We need to consider $\tau_{\textsf{\tiny FLL}}<t_{e}<\tau_{\textsf{\tiny r}}$. 

\begin{figure}
\centering
\includegraphics[width=1.0\columnwidth]{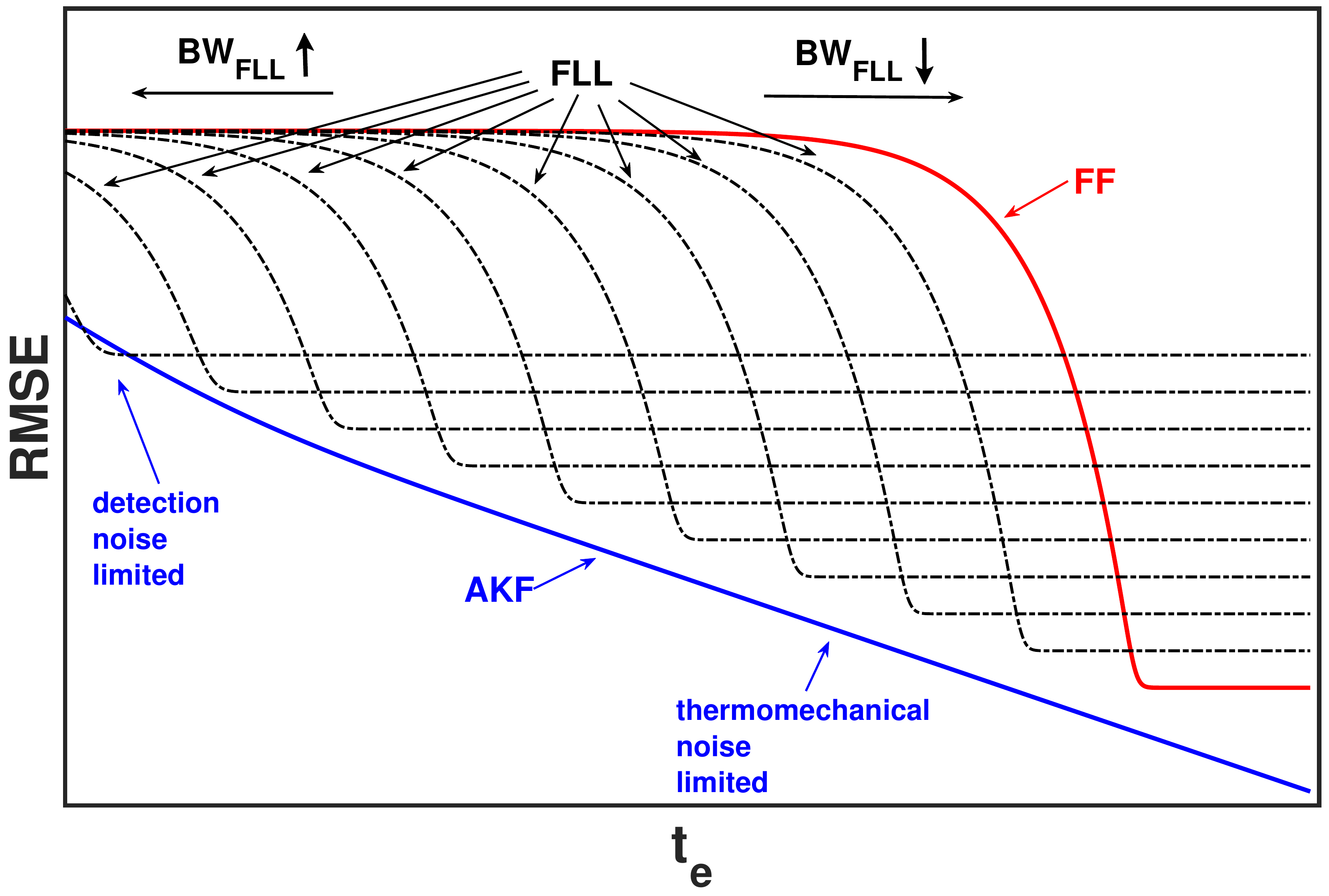}
\caption{\textsf{\small RMSE} vs elapsed time for {\sc FF}, {\sc FLL}, {\sc AKF}. Plot is log-log with both \textsf{\small RMSE} and $t_{e}$ unitless and normalized.} 
\label{fig:mse}
\end{figure}

For {\sc FLL} and {\sc SSO}, one can choose the detection bandwidth in the range from the resonator linewidth, $\omega_{\textsf{\tiny r}}/{Q}$, all the way to the resonance frequency $\omega_{\textsf{\tiny r}}$. 
In fact, when the detection bandwidth is set to the linewidth of the resonator, {\sc FLL} becomes equivalent to the {\sc FF} scheme. 
For a given detection bandwidth, \textsf{\small MSE} for these schemes reaches the minimum asymptotic value, \ie the variance in \eqref{eqn:varFFFLLSSO}, when the bias becomes zero, which occurs approximately at 
\begin{equation} 
t_{e}^{m} \approx 10\,\tau_{\textsf{\tiny FLL}} = 10\,\tau_{\textsf{\tiny SSO}} = \frac{5}{\textsf{BW}_{\textsf{\tiny FLL}}} = \frac{5}{\textsf{BW}_{\textsf{\tiny SSO}}}
\end{equation}
In order to compare {\sc AKF} with  {\sc FLL} and {\sc SSO}, we evaluate the $t_{e}$ dependent  \textsf{\small MSE} of {\sc AKF} at the above $t_{e}^{m}$. Here, we assume that $t_{e}^{m}> t_{e}^{*}$, where $t_{e}^{*}$ is given in \eqref{eqn:asympintersectAKF}. That is, we are assuming that $t_{e}^{m}$ falls into the thermomechanical noise limited regime for {\sc AKF}. The  \textsf{\small MSE} for {\sc AKF} can be found by evaluating \eqref{eqn:AKFvarhiasymp} at $t_{e}^{m}$ as follows
\begin{equation}
\textsf{\small MSE}^{\textsf{\tiny AKF}}\pqty{t_{e}^{m}} = \frac{\cS_{y_{\textsf{\tiny th}}}\,\textsf{BW}_{\textsf{\tiny FLL}}}{5} = \frac{\cS_{y_{\textsf{\tiny th}}}\,\textsf{BW}_{\textsf{\tiny SSO}}}{5}
\label{eqn:AKFMSEattm}
\end{equation}  
Then,
\begin{equation}
\begin{aligned}
\frac{\textsf{\small MSE}^{\textsf{\tiny FLL}}}{\textsf{\small MSE}^{\textsf{\tiny AKF}}}\pqty{t_{e}^{m}} = \frac{\textsf{\small MSE}^{\textsf{\tiny SSO}}}{\textsf{\small MSE}^{\textsf{\tiny AKF}}}\pqty{t_{e}^{m}} = 5\,\pqty{1+{\cal
        K}_{\textsf{\tiny d}}^2} 
\end{aligned}
\label{eqn:AKFFLLSCOMSErat}
\end{equation} 
Thus, \textsf{\small MSE} for {\sc AKF} is smaller than the minimum that can be achieved by fixed-bandwidth {\sc FLL} and {\sc SSO} at the same elapsed time after the event. The impact of detection noise on {\sc AKF} can be minimized, and in fact, almost eliminated, since the detection bandwidth $\textsf{BW}_{\textsf{\tiny L}}$ can be reduced all the way down to the resonator linewidth. On the other hand, for {\sc FLL} and {\sc SSO}, detection bandwidth $\textsf{BW}_{\textsf{\tiny FLL}}$ and $\textsf{BW}_{\textsf{\tiny SSO}}$ determine the response speed and can not be reduced with only noise filtering considerations. Furthermore, {\sc AKF}, featuring an adaptive bandwidth, continually improves the frequency shift estimate quality beyond the \textsf{\small MSE} value at  $t_{e}^{m}$. On the other hand, \textsf{\small MSE} for {\sc FLL} and {\sc SSO} has a lower bound that is reached at $t_{e}^{m}$ for a given bandwidth. A graphical illustration of \textsf{\small RMSE} versus elapsed time following event detection, is provided in Fig.~\ref{fig:mse} for {\sc FF}, fixed-bandwidth {\sc FLL}, and {\sc AKF}. With this illustration, one can observe that the {\sc AKF} scheme conceptually behaves as an {\sc FLL} with a time-varying, adaptive-bandwidth that is dramatically increased after event detection and gradually reduced thereafter. However, due to the optimality properties associated with {\sc AKF}, it outperforms such a conceptual adaptive {\sc FLL} as quantified by  \eqref{eqn:AKFFLLSCOMSErat}. We should also note that turning the adaptive {\sc FLL} concept into a practical scheme would amount to a nontrivial engineering endeavor, a pointless exercise in our opinion given the optimality and generality of the {\sc AKF} scheme.    
  
We note that we used the standard definition of variance in \eqref{eqn:mseakf}, in contrast with the use of Allan variance in~\cite{demir2020sensortradeoffs}. This is due to the fact that Allan variance is not appropriate for characterizing the accuracy of the frequency shift estimates provided by the {\sc AKF} based scheme. As discussed above, {\sc AKF} provides estimates with a time-varying variance, conceptually corresponding to a time-varying spectral density for fractional frequency fluctuations, \ie $\cS_{y_{\textsf{\tiny r}}}\pqty{\omega}$ depends on $t_{e}$. This would make Allan variance in \cite[eqn.~75]{demir2020sensortradeoffs} not only a function of the averaging time $\tau$, but also a function of the elapsed time $t_{e}$, rendering it confusing and uninterpretable. The factor in \cite[eqn.~75]{demir2020sensortradeoffs} that multiplies  $\cS_{y_{\textsf{\tiny r}}}\pqty{\omega}$ represents an averaging operation (over a duration of $\tau$), that is not an integral part of the sensing scheme but simply dictated by the definition of Allan variance. On the other hand, the time-varying variance for {\sc AKF} (decreasing function of elapsed time $t_{e}$) represents an inherent noise averaging operation that is an integral part of the sensing scheme itself. For {\sc FF}, {\sc FLL}, and {\sc SSO}, there is no inherent noise averaging, resulting in a constant standard variance. Thus, comparisons of {\sc FF}, {\sc FLL}, and {\sc SSO} with {\sc AKF} based on an error definition that mixes the two different kinds of averaging operation would become confusing and in fact incorrect.           
  
\section{Practical considerations for Adaptive Kalman Filter based resonator tracking}  
\label{sec:akfpractical} 

The {\sc AKF} based resonator tracking scheme can be implemented as an add-on to the {\sc FF} scheme in Fig.~\ref{fig:openloop}. The frequency shift output of the {\sc FF} scheme is simply scaled by the resonance frequency in order to obtain the fractional frequency shift, which would then serve as the observation $y_{\textsf{\tiny ro}}$ in \eqref{eqn:resonatorfffracfreq}. Thus, {\sc AKF} based resonator tracking amounts to processing the {\sc FF} scheme output. 
This can be conveniently done in the sampled-data domain with a digital implementation. {\sc AKF} can be implemented as an add-on software module in {\sc FPGA/DSP} based lock-in setups, either for realtime/online operation or offline post-processing. The simple but accurate resonator model with very low complexity, shown in Fig.~\ref{fig:openloopphase}, is the key enabler for a robust {\sc AKF} based resonator tracking technique that can operate in realtime. It should be noted that the simple resonator model requires the values of two resonator parameters, the resonance frequency $\omega_{\textsf{\tiny r}}$ and the quality factor $Q$, which can be measured in a pre-characterization step, and occasionally updated during {\sc AKF} operation.    

When {\sc AKF} is implemented as an add-on at the output of the {\sc FF} scheme as described above, the disadvantages of the open-loop {\sc FF} scheme relating to resonance frequency drifts, usually due to slow thermal phenomena, are inherited~\cite{demir2020sensortradeoffs}. However, this problem can be solved easily. The most straightforward and simplest approach would be to implement an {\em add-on} {\sc FLL} with a loop bandwidth that is (much) smaller than the linewidth of the resonator, that is, a very slow {\sc FLL}. The purpose of the {\sc FLL} here is not to sense and track the resonance frequency shifts at fast time scales due to events of interest, but simply to keep the drive of the resonator close to the resonance over long time scales, compensating for resonance frequency drifts due to nonidealities as well as possibly accumulating frequency shifts due to events of interest. This {\sc FLL} would operate on the (very) slow speed, (very) high accuracy extreme of the speed-accuracy trade-off characteristics. Since this {\sc FLL} is very slow, the output of the demodulator will essentially reflect the open-loop, feedback-free resonator response at short time scales up to and exceeding the resonator time constant. The filter in the demodulator in this case has a bandwidth set to the linewidth of the resonator, but the loop bandwidth for the {\sc FLL} can be set to a value that is (much) smaller than the linewidth by choosing the controller parameters accordingly as described in~\cite{demir2020sensortradeoffs}. While the {\sc FLL} is making slow corrections to the drive frequency to keep the resonator operating around its resonance, the {\sc AKF} will detect, predict, estimate and track the faster variations in resonance frequency due to events of interest. Alternatively, instead of using an {\sc FLL} in order to compensate for drifts and accumulated frequency shifts, one can directly use the output of the {\sc AKF} itself in order to continually, and at a faster time scale, update the drive frequency for the resonator.

\section{Conclusions}
\label{sec:concl}

We have shown that there is still room for improvement in nanomechanical resonant sensors with sensor architecture innovations. Pareto optimality with respect to speed and accuracy can be achieved by using an adaptive Kalman filtering based scheme that is founded on well established techniques in detection and estimation theory. In this paper, we have assumed an event where the resonance frequency experiences a sudden shift that persists, \eg when a particle falls on and sticks to a {\sc NEMS} resonator.  However, the {\sc AKF} technique presented can be easily modified to detect and track events where the resonance frequency experiences a sudden but transient shift: Any type of event with a resonance frequency shift signature that can be modeled as the impulse response of a simple linear system can be easily incorporated into the {\sc AKF} scheme. In the case considered in this paper, events have a frequency shift signature that is modeled as the impulse response of an ideal integrator, {\it i.e.,} a step change. Replacing the integrator with a simple first-order filter with a prescribed time constant may be used as the model for a transient event signature. Even if the time constant of the event is much smaller than the resonator response time, {\sc AKF} will still be able to detect and extract the magnitude of the root cause of the transient event. Finally, the proposed {\sc AKF} approach is certainly not restricted to nanomechanical resonant sensors, it can also be used for optical and microwave resonant sensors. 

\section*{Acknowledgment}
The author would like to thank Selim Hanay and Alper Erdogan for helpful discussions.   

\bibliographystyle{IEEEtran}
\bibliography{ref}

\begin{IEEEbiography}[{\includegraphics[width=1.0in,height=1.25in,clip,keepaspectratio]{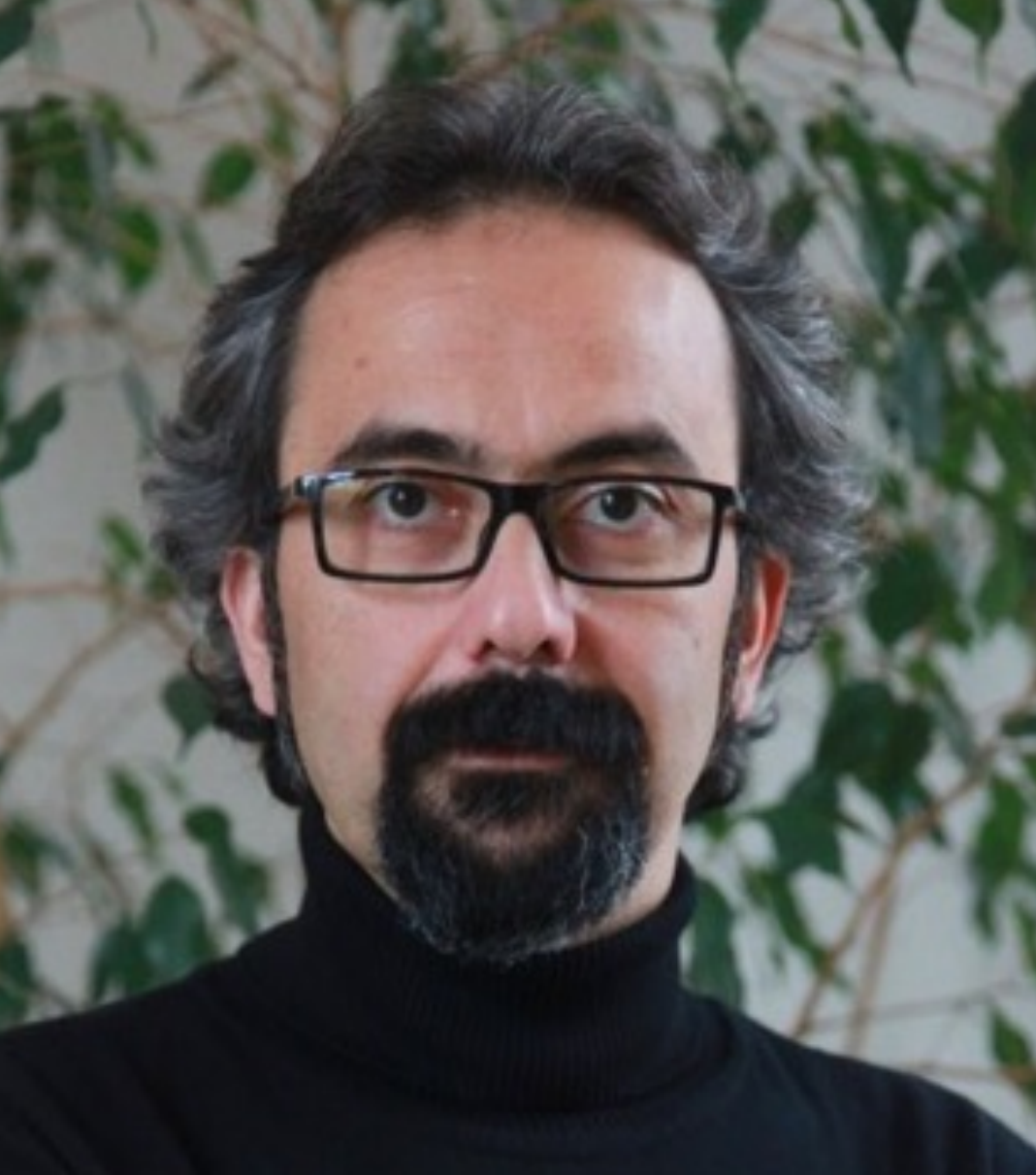}}]
{Alper Demir}
received a B.S. from Bilkent University, Ankara,
Turkey, in 1991, and a Ph.D. from the University of
California, Berkeley, CA, USA, in 1997.  He was a member of technical staff at
Bell Laboratories, Murray Hill, NJ, USA. He has been with 
Ko\c{c} University, Istanbul, Turkey, since 2002.  He received 
the 2002 Best of ICCAD Award, the 2003/2014 IEEE/ACM
William J. McCalla ICCAD best paper awards, and the 2004 IEEE
Guillemin-Cauer Award. He is an IEEE Fellow.  
\end{IEEEbiography}

\end{document}